\documentclass[a4paper]{article}

\usepackage{INTERSPEECH2021, url, multirow, amsmath, pifont, amssymb, makecell}

\newcommand{\cmark}{\text{\ding{51}}}
\newcommand{\xmark}{\text{\ding{55}}}

\title{Improving weakly supervised sound event detection with self-supervised auxiliary tasks}
\name{Soham Deshmukh$^{1}\thanks{The work was done at Carnegie Mellon University}$, Bhiksha Raj$^2$, Rita Singh$^2$}
\address{
  $^1$Microsoft, $^2$Carnegie Mellon University}
\email{sdeshmukh@andrew.cmu.edu, bhiksha@cs.cmu.edu, rsingh@cs.cmu.edu}

\begin{document}

\maketitle
\begin{abstract}
While multitask and transfer learning has shown to improve the performance of neural networks in limited data settings, they require pretraining of the model on large datasets beforehand. In this paper, we focus on improving the performance of weakly supervised sound event detection in low data and noisy settings simultaneously without requiring any pretraining task. To that extent, we propose a shared encoder architecture with sound event detection as a primary task and an additional secondary decoder for a self-supervised auxiliary task. We empirically evaluate the proposed framework for weakly supervised sound event detection on a remix dataset of the DCASE 2019 task 1 acoustic scene data with DCASE 2018 Task 2 sounds event data under 0, 10 and 20 dB SNR. To ensure we retain the localisation information of multiple sound events, we propose a two-step attention pooling mechanism that provides a time-frequency localisation of multiple audio events in the clip. The proposed framework with two-step attention outperforms existing benchmark models by 22.3 \%, 12.8 \%, 5.9 \% on 0, 10 and 20 dB SNR respectively. We carry out an ablation study to determine the contribution of the auxiliary task and two-step attention pooling to the SED performance improvement.\footnote{The code is publicly released.}.
\end{abstract}

\section{Introduction}
\label{sec:introduction}
Sound Event Detection (SED) aims to determine the presence, nature and temporal location of sound events in audio signals. Many SED algorithms rely on strongly labelled data \cite{SLD_1, SLD_2, SLD_3} for training to perform accurate event detection and localisation. However, producing strongly labelled data for SED is quite expensive in terms of the expertise, time and human resources required for the annotation. This has led to the creation of weakly labelled sound event detection dataset like Audioset \cite{Audioset} which contains audio clip level annotations without the corresponding onset and offset times of the audio events.

The weakly supervised sound event detection was first formulated as a Multiple-Instance Learning (MIL) problem \cite{MIL, Anurag_WLD} with the recent emergence of Neural MIL. In Neural MIL, the first half of the network (segmentation network) produces temporal predictions which are then aggregated by the second half of the network (classification network) usually a pooling operator to produce audio clip level predictions. 
The benefit of such formulation is, along with detecting audio events in the clip, it provides insight into time level localisation of those sound events in the audio clip. Since then, recent works have focused on improving the model architecture of the segmentation network \cite{Florian, DeepCNN_anurag, gcnn} and developing better pooling methods \cite{gwrp, framecnn, adaptive_pooling, gaussianfilter, atrous}. However, few works have focused on how sound event detection models perform in either limited data or noisy settings let alone in both of them.


The noisy data also affects the training of networks for sound event detection.  Specifically, the deep CNN architectures \cite{vgg, resnet} currently used to provide benchmark performance for different speech and audio tasks \cite{largescale_audioclassification} require large labelled clean datasets to train on and when considered in a noisy environment the performance is known to deteriorate \cite{gwrp}. 
The two general learning strategies used as solutions are transfer learning and multitask learning which were recently utilised for sound event detection \cite{Xue2020SoundEL, Imoto2020SoundED, andrew}.
However, in the multitask learning setup, it's assumed you have richly annotated labels for all the tasks. We investigate a counterpart of this where only weak labels are available without any labels for the secondary task. For this setting,  we propose a self-supervised auxiliary task that will be jointly trained with the primary task of sound event detection. The auxiliary task is chosen to be the reconstruction of log Mel spectrogram of audio and we show how the auxiliary task denoises internal representations and improves network performance in noisy settings.

In all, in this paper, we address the challenge of training sound event detection models in noisy (domestic or environmental) and limited data settings. To that effort, we make two-fold contributions. First, identify appropriate self-supervised auxiliary task for sound event detection in noisy settings and demonstrate performance benefits to the same. Second, develop a two step attention pooling mechanism that improves time-frequency localisation of audio events and indirectly improves sound event detection performance in noisy settings. We perform all the experiments on a standard noisy sound event detection dataset remix \cite{gwrp} and release the code publicly.

\begin{figure*}
    \centering
    \includegraphics[width=6.5in]{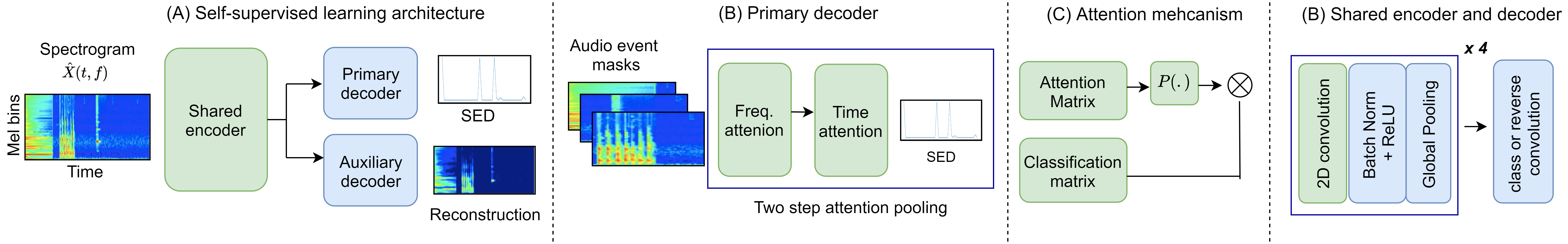}
    \caption{Our proposed self-supervised learning assisted framework for weakly supervised sound event detection. (A) The general architecture with shared encoder and multiple decoder branches. Shared encoder, primary decoder, auxiliary decoder is represented by $g$, $g_2$, $g_4$ respectively (B) shows the two step attention pooling function used for primary decoder. (C) The attention mechanism used for frequency and time attention in two step attention pooling along different axis. (D) The CNN architecture used for shared encoder and auxiliary decoder. The last layer is either class or reverse convolution for encoder and decoder respectively.}
    \label{fig:MTL_setup}
\end{figure*}

\section{Related work}
\label{sec:related_work}
A prominent recent work \cite{gwrp} analysed the performance of different model architectures (segmentation network and pooling functions) under different Signal to Noise Ratio (SNR) for sound event detection and localisation. The paper showed that the segmentation network of type `VGG-like' CNN performed best for audio tagging and variability in performance resulted from the choice of pooling methods with not a clear winning pooling method across different SNR. Specifically, Global Attention Pooling outperformed other pooling methods on some SNR and metrics, while Global Weighted Rank Pooling (GWRP) results in the best performance on others. Still, the work on sound event detection performance in limited data and noisy settings is sparse.

Though the various type of multitask learning methods have been greatly explored for vision and natural language processing (NLP) tasks \cite{mtl_survey}, it has not been utilised by the audio community. Most of the works in multitask learning for SED focus on jointly training SED with another strongly labelled task like Sound Source Localisation (SSL) \cite{Xue2020SoundEL} or Acoustic Scene Classification (ASC) \cite{Imoto2020SoundED, Tonami2019JointAO, Bear2019TowardsJS}. A combination of multitask learning and self-supervised learning is shown to improve performance on speech and audio tasks \cite{andrew}. However, the work uses large scale speech datasets like Librispeech \cite{librispeech} as pretasks to pretrain the networks using self-supervised learning and does not analyse the effect of noise (domestic or environmental) on sound event detection performance.

\section{Methodology}
\label{sec:methodology}
This section contains the details of the proposed approach for SED, segmentation mapping network $g_1$, classification mapping network $g_2$, and the auxiliary time-frequency reconstruction auxiliary task. The architecture is depicted in figure \ref{fig:MTL_setup}

\subsection{Self-supervised Learning formulation for SED} \label{section:proposedsolution_formulation}
Let the raw audio be represented by $X = \{{x_i}\}_{i = 1}^T$ where each $x_i \in \mathbb{R}$ is a frame in the audio clip. We extract time-frequency features for each audio, let them be represented by $\hat{X} = \{{\hat{x}_i}\}_{i = 1}^T$ where each $\hat{x_i} \in \mathbb{R}^d$, $d \in \mathbb{Z}$ corresponds to frame in the audio clip. In practice, d are the number of mel bins obtained after computing the spectrogram. As per MIL formulation, we can represent each sample in dataset as a bag $B_j = (\{\hat{x}_i\}_{i = 1}^T, y)\vert_{j = 0}^N$ where $y \in \mathbb{R}^C$ is the weak label, N are the number of samples and C are the number of audio events. The primary task in our self-supervised framework is SED. The segmentation mapping $g_1(.)$ of SED also acts as a shared encoder for the auxiliary task. The shared encoder maps the feature set $\{{\hat{x}_i}\}_{i = 1}^T$ to $Z = \{z_i\}_{i = 1}^T$ where $z_i \in \mathbb{R}^{C \times F \times T}$. The second part of SED task is network which classifies $\{z_i\}_{i = 1}^T$ to $P = \{ p_i \}_{i = 1}^C$ where $P \in \mathbb{R}^C$. The network learns a mapping $g_2$ which maps each audio events time-frequency segmentation to corresponding presence probabilities of $c^{th}$ event known as $p_k$
\begin{equation}
    g_1 : \hat{X} \mapsto Z \quad\quad g_2 : Z \mapsto P
\end{equation}

The auxiliary self-supervised task chosen needs to help in learning robust representations which generalise to noisy settings without requiring additional labels. This will impact not only the learned internal representation but also downstream sound event detection and localisation performance. Inorder to achieve that we choose auxiliary task as reconstruction of extracted time-frequency features for audio. By having time-frequency reconstruction auxiliary task we hypothesise the network will learn representations which retain audio event information better \cite{denoise1, denoise2}. We use an auto-encoder structure for reconstruction where the encoder is shared with the primary task of SED. If $g_3(.)$ is encoder mapping for reconstruction task, we now represent $g_1(.) = g_3(.) = g(.)$ as the shared segmentation mapping function. The second part of auxiliary task, is a decoder network which learns a mapping $g_4$ such that $g_4: Z \mapsto \bar{X}$ where $\bar{X}$ is the reconstructed time-frequency representation. Specifically $\{\bar{x_i}\}_{i = 1}^T = g_4(\{z_i\}_{i = 1}^T)$. Here the learned mapping function $g_4(.)$ should satisfy: 
\begin{equation}
    g_4^{-1}(g(.)) = g^{-1}(g_4(.)) = I
\end{equation}
To learn the function mappings satisfying primary SED task, let the objective function be $\mathcal{L}_1$. To enforce the constraint of auxiliary task, let the objective function be $\mathcal{L}_2$ where the aim is to minimise the difference between T-F representation $\{\hat{x_i}\}_{i = 1}^T$ and predicted time-frequency representation $\{\bar{x_i}\}_{i = 1}^T$ of audio clip. If the learnable parameters are W = $[w, w_2, w_4]$ and $w, w_2, w_4$ corresponding to $g(.), g_2(.), g_4(.)$ respectively, then the optimisation problem can be framed in terms of these weights W over all data points as:
\begin{equation}
    \underset{W}{\text{min}}\; \mathcal{L}_1(P, y \vert w, w_4) + \alpha \mathcal{L}_2(\{\bar{x_i}\}_{i = 1}^T, \{\hat{x_i}\}_{i = 1}^T \vert w, w_2)
\end{equation}
The parameter alpha ($\alpha$) accounts for scale difference between losses $\mathcal{L}_1$ and $\mathcal{L}_2$. It helps in adjusting the contribution of auxiliary task relative to the primary task in learning weights.

\subsection{Shared encoder and auxiliary task decoder network} \label{section:Shared Segmentation amand Decoder Network}
The segmentation mapping function (shared encoder) converts the time-frequency audio input into a T-F representation for each of the audio events. The time-frequency feature extracted for audio here is log Mel spectrogram as it has shown to provide better performance \cite{environmental_sound_classification, esc_deepcnn, largescale_audioclassification}. We choose a CNN based architecture similar to `VGG-like' \cite{gwrp} for both shared encoder and auxiliary task decoder. 
The shared encoder has CNN based network consists of 8 blocks of 2D Convolution, BatchNorm and ReLU with an Average Pool after every 2 blocks. Having a common encoder helps the network to learn a shared representation by exploiting the similarity across SED and T-F reconstruction and enables the network to generalise better on our original task. We use a hard parameter sharing framework to reduce the risk of overfitting \cite{mtl_informationtheory} to limited samples. 


\begin{table*}[!t]
\footnotesize
\centering
\caption{Weakly supervised sound event detection performance across different SNR} 
\label{table:sed}
\begin{tabular}{|c|c|c|c|c|c|c|c|c|c|c|c|}
\hline
\multicolumn{3}{|c|}{Network} & \multicolumn{3}{c|}{SNR 20 dB} & \multicolumn{3}{c|}{SNR 10 dB} & \multicolumn{3}{c|}{SNR 0 dB} \\
encoder & pooling & \multicolumn{1}{l|}{aux.} & micro-p & macro-p & AUC & micro-p & macro-p & AUC & micro-P & macro-p & AUC \\ \hline
VGGish & GAP & \xmark & 0.5067 & 0.6127 & 0.9338 & 0.4291 & 0.5390 & 0.9144 & 0.3295 & 0.4093 & 0.8694 \\ \hline
VGGish & GMP & \xmark & 0.5390 & 0.5186 & 0.8497 & 0.5263 & 0.5023 & 0.8422 & 0.4640 & 0.4441 & 0.8189 \\ \hline
VGGish & GWRP & \xmark & 0.7018 & 0.7522 & 0.9362 & 0.6538 & 0.7129 & 0.9265 & 0.5285 & 0.6084 & 0.8985 \\ \hline
VGGish (dil.) & AP & \xmark & 0.7391 & 0.7586 & 0.9279 & 0.6740 & 0.7404 & 0.9211 & 0.5714 & 0.6341 & 0.9014 \\ \hline
VGGish & 2AP & \cmark & \textbf{0.7829} & \textbf{0.7645} & \textbf{0.9390} & \textbf{0.7603} & \textbf{0.7486} & \textbf{0.9343} & \textbf{0.6986} & \textbf{0.6892} & \textbf{0.9177} \\ \hline
\end{tabular}
\end{table*}
The decoder of the auxiliary-task takes $Z = \{z_i\}_{i = 1}^T$ as input and reconstructs it to $\{\bar{x_i}\}_{i = 1}^T$. The decoder consists of CNN based network for combining the intermediate time-frequency representations obtained for each audio event to an audio level time-frequency representation. The architecture closely follows the common encoder structure in reverse order consisting of 8 blocks of 2D Convolution, BatchNorm and ReLU with Average Pool after every two blocks with a decreasing number of filters.

\subsection{Primary decoder \label{section:classification_network}}
The primary decoder is not CNN based, instead, it is a pooling operator to satisfy MIL formulation. The choice of pooling operator has a significant performance effect on both the SED and each audio events intermediate time-frequency representation obtained. Global max pooling and global average pooling results in underestimate and overestimate the audio event's temporal presence respectively, and to overcome this problem dynamic poolings were proposed \cite{gwrp, adaptive_pooling, atrous}. However, the developed pooling mechanisms still lacks the granularity in temporal predictions and does not provide frequency localisation which might be used to further disambiguate sound events. Also, the standard attention pooling \cite{atrous} is known to be unstable with cross-entropy usually used for multi-class setup in practice.

We propose a two-step attention pooling mechanism to covert each audio events segmentation maps $\{z_i\}_{i = 1}^T$ into audio level predictions $P$. The first step in the two step attention pooling takes $Z = \{z_i\}_{i = 1}^T$ as input. This undergoes two independent learned linear transformation to produce classification and attention output respectively. The attention output is squashed to ensure its valid probability distribution. Mathematically, the attention output $Z_{a_1}$ and classification output $Z_{c_1}$ are:
\begin{equation}
    Z_{a_1} = \frac{e^{\sigma(ZW_{a_1}^T + b_{a_1})}}{\sum_{i = 1}^F e^{\sigma(ZW_{a_1}^T + b_{a_1})}} \quad Z_{c_1} = (ZW_{c_1}^T + b_{c_1})
\end{equation}
This is followed by a weighted combination of classification output $Z_{c_1}$ by attention weights $Z_{a_1}$:
\begin{equation}
    Z_{p_1} = \sum_{i = 0}^F Z_{c_1} \cdot Z_{a_1}
\end{equation}
The time level attention is similar to frequency (first step) attention except it operates along time axis:
\begin{equation}
    Z_{a_2} = \frac{e^{\sigma(Z_{p_1}W_{a_2}^T + b_{a_2})}}{\sum_{t = 1}^T e^{\sigma(Z_{p_1}W_{a_2}^T + b_{a_2})}} \quad Z_{p_1} = (ZW_{c_2}^T + b_{c_2})
\end{equation}
\begin{equation}
    Z_{p_2} = \sum_{t = 0}^T Z_{c_2} \cdot Z_{a_2}
\end{equation}
where $Z_{p_2} \in [0,1]$ and denotes the presence probability of each sound event in the audio clip. Figure \ref{fig:MTL_setup} subsection c, provides an overview of a single attention step. 
In relation to figure, $Z_{a}$, $Z_{c}$, $Z_{p}$ are the outputs after attention matrix, classification matrix and $P(.)$ respectively in the first stage and second stage depending on subscript. By breaking the attention into two steps, it makes the pooling more interpretable by answering the questions of what frequency bins and what time steps contributes to which audio events by visualising normalised attention weights $Z_{a_1}, Z_{a_2}$ and output $Z_{p_1}, Z_{p_2}$. Also, the sigmoid ($\sigma$) ensures the attention output stays between 0 to 1 and avoids unstable training for multilabel training with cross-entropy in practice.


\section{Experiments} \label{section:experiments}
\subsection{Dataset} \label{section:experiments_dataset}
To study the effect of noise in limited data settings, we form a noisy dataset by mixing DCASE 2019 Task 1 of Acoustic scene classification \cite{dcase1} and DCASE 2018 Task 2 of General purpose Audio tagging \cite{dcase2}. The DCASE 2019 Task 1 provides background sounds (noise) recorded from a variety of real world scenes in which the sounds from DCASE 2019 Task 2 are randomly embedded \cite{gwrp}. To ensure the noise conditions are natural, diverse and challenging, we use the new DCASE 2019 Task 1 instead of DCASE 2018 as used in \cite{gwrp}. The 2019 variant extends the TUT Urban Acoustic Scenes 2018 with the other 6 cities to a total of 12 large European cities. This results in 32000 audio clips with 8000 audio clips for each 20,10,0 dB SNR where each audio clip is of 10 secs with background noise and three random audio events (out of total 41) in it.

\subsection{Set up} \label{section:set up}
The raw data is converted to time-frequency representation by applying FFT with a window size of 2048 and an overlap of 1024 between windows. This is followed by applying Mel filter banks with 64 bands and converting them to log scale to obtain log Mel spectrogram. The network architecture used is described in section \ref{section:Shared Segmentation amand Decoder Network}. 
The entire network is trained end-to-end with a batch size of 24 and learning of 1e-3 using Adam optimiser \cite{adam}. The code and setup is publicly released\footnote{\url{https://github.com/soham97/MTL_Weakly_labelled_audio_data}}.

\section{Results} \label{section:results}

\subsection{Sound event detection} \label{section:sound event detection}
We evaluate our self-supervision assisted architecture and pooling method against different baselines, benchmark architectures and pooling methods \cite{atrous, gwrp}. Table \ref{table:sed} shows weakly supervised sound event detection performance across different SNR of 20,10, and 0 dB. The important evaluation metric here under consideration is micro precision (micro-p), as it uses global counts of true positives, false negatives and false positives for metric computation against macro precision which does simple unweighted averaging disregarding class-imbalance. The VGGish (dil.) encoder here indicates VGGish architecture but with dilated/atrous convolutions known to provide benchmark performance for sound event detection, \cite{atrous}. The VGGish encoder with reconstruction based auxiliary task and two step attention pooling outperforms the existing benchmark of atrous attention pooling \cite{atrous} on SNR 20, 10 and 0 dB by 5.9\%, 12.8\% and 22.3\% respectively. Apart from improving performance, by breaking the attention into two steps, it allows for the intermediate use of sigmoid which helps in ensuring the outputs don't overflow above 1 during training.

\begin{table}[t]
\footnotesize
\centering
\caption{Ablation study to determine auxiliary task contribution} 
\label{table:ablation_study}
\begin{tabular}{|c|c|c|c|}
\hline
auxiliary task & SNR 20 dB & SNR 10 dB & SNR 0 dB \\ \hline
$\alpha$ = 0.0 & 0.7772 & 0.7430 & 0.6937 \\ \hline
$\alpha$ = 0.001 & \textbf{0.7829} & \textbf{0.7603} & \textbf{0.6986} \\ \hline
$\alpha$ = 0.1 & 0.7637 & 0.7428 & 0.6792 \\ \hline
\end{tabular}
\end{table}

\subsection{Ablation study for auxiliary task contribution} \label{section:ablation study}

We perform an ablation study to determine the contribution of reconstruction auxiliary task and two step attention pooling towards the total performance improvement. As described in Section \ref{section:proposedsolution_formulation}, the total loss is:
\begin{align}
    \mathcal{L} = \mathcal{L}_1(P, y \vert w, w_4) + \alpha \mathcal{L}_2(\{\bar{x_i}\}_{i = 1}^T, \{\hat{x_i}\}_{i = 1}^T \vert w, w_2)
\end{align}
By changing the value of $\alpha$ before training, we can adjust the contribution of the auxiliary task to primary sound event detection. When $\alpha = 0.0$, the network has no contribution from the reconstruction auxiliary task during training and it can be used to evaluate the performance of two step attention pooling. In terms of micro-precision, the two step attention pooling outperforms existing benchmark of atrous AP (row 4) from table \ref{table:sed} on SNR 20, 10 and 0 dB by 5.2\%, 10.2\% and 21.4\% respectively. By adding the auxiliary task contribution with a relative weightage of $\alpha = 0.001$, an additional improvement of 0.7\%, 2.3\% and 0.7\% is observed. This indicates that two step attention has a prominent contribution in improving the performance of sound event detection in limited data and noisy settings, with additional performance gains from the auxiliary task. When $\alpha$ is increased to 0.01, the performance compared to $\alpha = 0.001$ is decreased. This suggests that the auxiliary task's loss contribution starts to overpower the primary SED task's loss contribution rather than improving generalisation. 

\begin{table}[h]
\footnotesize 
\centering
\caption{Two top and worst performing sound events- SNR 0 dB} 
\label{table:class_sed}
\begin{tabular}{|l|l|l|l|l|l|}
\hline
model & aux.& bus & cowbell & gong & meow \\ \hline
Atrous + AP & \xmark & 0.2 & 0.781 & \textbf{0.692} & \textbf{0.583}\\ \hline
VGGish + 2AP & \xmark & 0.572 & 0.921 & 0.643 & 0.483\\ \hline
VGGish + 2AP & \cmark & \textbf{0.627} & \textbf{0.94} & 0.663 & 0.532\\ \hline
\end{tabular}
\end{table}

\subsection{SED performance on specific audio events}
For almost all audio events, our proposed architectures have the best precision scores against GMP, GAP, GWRP, Atrous across all SNR = 0, 10, 20. Particularly, for audio events like `Bass drum', `bus', `double bass', `cowbell' the architecture outperforms other models by a large margin as shown in table \ref{table:class_sed}. However, the proposed model struggles in audio events like `gong',`chime' and `meow' where the attention pooling with dilated convolution encoder performs better \cite{atrous}. This indicates using atrous or dilated convolutions helps in detecting audio events whose energy is spread wide in the temporal domain. This can be incorporated into our current architecture by replacing the linear convolutions in the shared encoder with dilated convolutions. Further analysis and event-specific results are available in the long version of paper \footnote{https://arxiv.org/pdf/2008.07085.pdf} and skipped due to space constraints.

\begin{figure}
    \centering
    \includegraphics[width=3.1in]{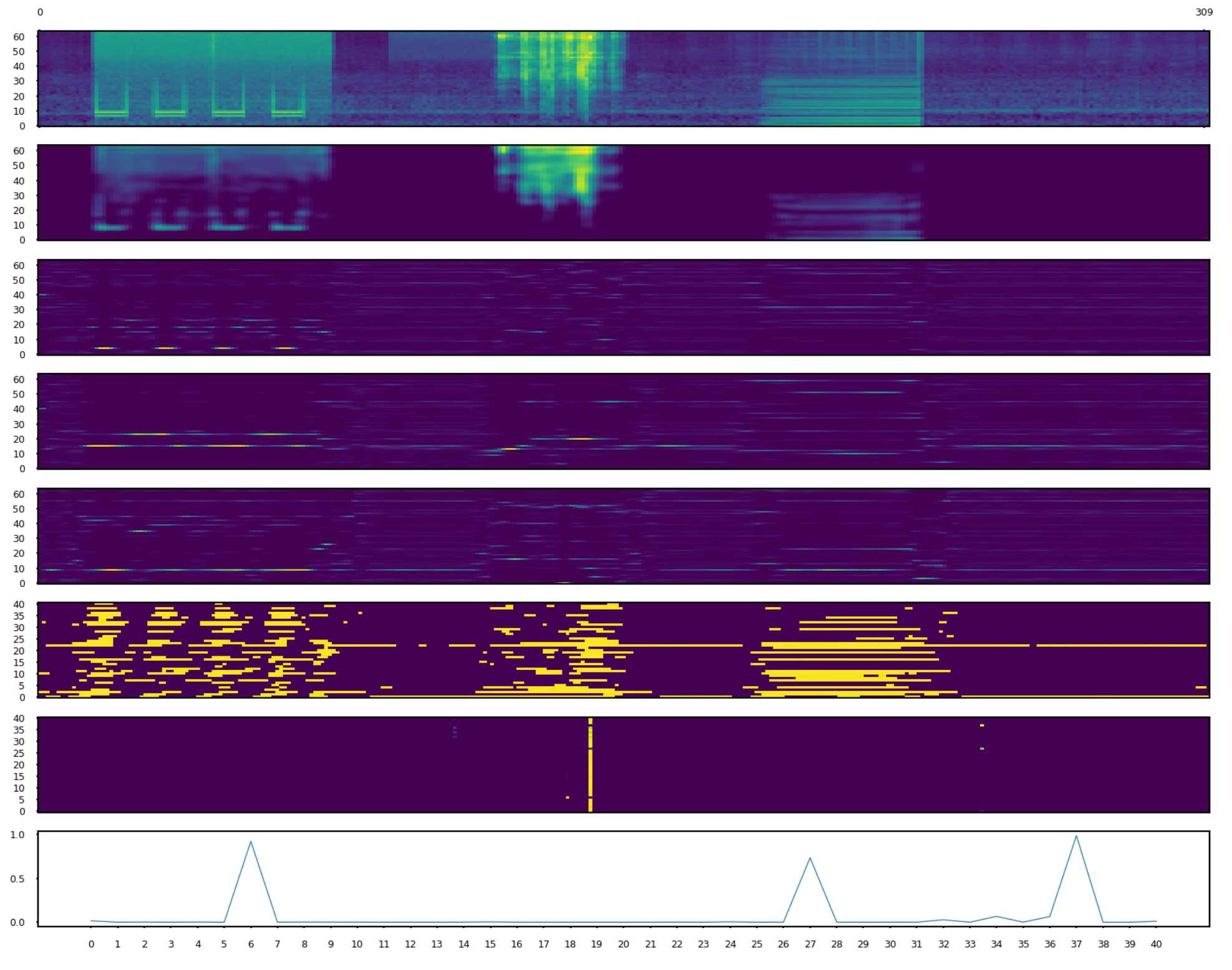}
    \caption{Visualisation of two step attention pooling and reconstruction decoder outputs. Subplot 1 depicts the scaled log Mel spectrogram of an audio clip. Subplot 2 is the output of the reconstruction auxiliary task. Subplots 3,4 and 5 are attention weights for the three most probable audio events in the audio clip. Subplot 6 is the output of the first step attention pooling. Subplot 7 and 8 is the attention weight and output of second step attention pooling respectively. The y-axis in subplot 1-4 corresponds to Mel-bins and sound events in subplot 5-6. The x-axis in subplot 1-7 corresponds to time and sound events in subplot 8}
    \label{fig:visualisation}
\end{figure}

\subsection{Interpretable visualisation of audio events\label{section:results_vis}}
Apart from improved performance, using two step attention pooling provides a way to localise each audio event present in the audio clip along with both the time and frequency axis. To illustrate this, we pick a random example with SNR 20 dB and show the end to end visualisation of the two step attention pooling mechanism in figure \ref{fig:visualisation}. The audio under consideration has three events occurring in it: telephone ringing, cello playing and cat meowing, with outdoor environmental background noise. Subplot 2 in figure \ref{fig:visualisation} depicts the reconstructed Mel spectrogram of the audio clip. From the subplot, we can see that the decoder is not only able to reconstruct the audio events clearly but it is also denoising the log Mel spectrogram retaining the key elements of three audio events. A future extension of work is to jointly train sound source separation along with weakly supervised SED by using the auxiliary task reconstruction output.

\section{Conclusions}
This paper proposes assisted self-supervised task for improving sound event detection in limited data and noisy settings. The architecture consists of sound event detection as a primary task with two-step attention pooling as a primary decoder and time-frequency representation reconstruction as an auxiliary task. We empirically evaluate the proposed framework for multi-label weakly supervised sound event detection, on a remix DCASE 2019 and 2018 dataset under 0, 10 and 20 dB SNR. The proposed self-supervised auxiliary task framework with two-step attention outperforms existing benchmark models by 22.3 \%, 12.8 \%, 5.9 \% on 0, 10 and 20 dB SNR respectively. The ablation study carried out indicates the majority of performance improvement is associated with two step attention pooling with secondary performance improvement from self-supervised auxiliary task. Furthermore, by using two step attention, we can easily visualise the sound event presence along both time-frequency axis. The code is publicly released. 

\bibliographystyle{IEEEtran}

\bibliography{main}


\end{document}